\documentclass[letterpaper, 10pt, conference]{ieeeconf}  
\IEEEoverridecommandlockouts

\usepackage{graphicx}       

\usepackage{amsfonts}
\DeclareMathSymbol{\shortminus}{\mathbin}{AMSa}{"39}
\let\labelindent\relax
\usepackage{enumitem}
\usepackage{multirow}
\usepackage{booktabs}
\usepackage{array} 

\usepackage{bm}
\usepackage{epsfig}
\usepackage{arydshln}
\usepackage{algpseudocode}
\usepackage{verbatim}
\usepackage{subfigure}
\usepackage{rotating}
\usepackage{color}
\usepackage{flushend} 
\usepackage{cite}

\usepackage{soul}
\usepackage{fancyhdr}
\usepackage{url}
\usepackage{makecell} 
\usepackage{mathtools}
\usepackage{caption}
\usepackage[para,online,flushleft]{threeparttable}

\usepackage{hyperref}       
 
\usepackage{nicefrac}       
\usepackage{microtype}      
\usepackage{xcolor}         

\usepackage{amsthm, amsmath, amssymb}
\usepackage{algorithm}

\newtheorem{definition}{Definition}
\newtheorem{problem}{Problem}
\newtheorem{theorem}{Theorem}
\newtheorem{remark}{Remark}

\usepackage{flushend}
 
\title{\bf Stability Verification for  Switched Systems   using \\
Neural Multiple Lyapunov Functions} 

\author{Junyue Huang, Shaoyuan Li and Xiang Yin
	\thanks{This work was supported by  the National Natural Science Foundation of China (62573291,62173226) and Science Center Program of National Natural Science Foundation of China under Grant 62188101.}
 \thanks{A preliminary version of this work has been accepted for presentation at the 64th IEEE Conference on Decision and Control (CDC 2025). Junyue Huang, Shaoyuan Li and Xiang Yin are with the School of Automation and Intelligent Sensing, Shanghai 200240, China.
	e-mail: \tt\small $\{$hjy-564904993, syli, yinxiang$\}$@sjtu.edu.cn}
}

\begin{document}
\maketitle

\begin{abstract}     
Stability analysis of switched systems, characterized by multiple operational modes and switching signals, is challenging due to their nonlinear dynamics.
While frameworks such as multiple Lyapunov functions (MLF) provide a foundation for analysis, their computational applicability is limited for systems without favorable structure.
This paper investigates stability analysis for switched systems under state-dependent switching conditions. We propose neural multiple Lyapunov functions (NMLF), a unified framework that combines the theoretical guarantees of MLF with the computational efficiency of neural Lyapunov functions (NLF). 
Our approach leverages a set of tailored loss functions and  a counter-example guided inductive synthesis (CEGIS) scheme
to train neural networks that rigorously satisfy MLF conditions. Through comprehensive simulations and theoretical analysis, we demonstrate NMLF's effectiveness and its potential for practical deployment in complex switched systems.
\end{abstract}

\section{Introduction}
  
Switched systems, composed of multiple subsystems and a switching signal that controls transitions between them, are widely used in engineering applications such as bipedal walking robots \cite{veer2020switched}, power electronics \cite{fribourg2013control}, and automotive control \cite{vargas2018switching}. A fundamental challenge in analyzing these systems lies in \emph{stability verification,} since the complex interaction between subsystem dynamics and switching mechanisms can potentially lead to unstable behavior.  Ensuring stability is crucial for reliable operation and for preventing performance degradation or system failure.
 
Over the past decades, numerous methodologies have been developed for stability verification of switched systems; see, e.g., \cite{ahmadi2014joint,wang2016stability,fiacchini2018stabilization,yang2024learning,yin2024formal}. A particularly prominent approach is the multiple Lyapunov functions (MLF) method \cite{branicky1998multiple,she2014discovering,long2017multiple}, which has emerged as an effective tool for stability analysis. The MLF framework employs subsystem-specific Lyapunov functions and establishes conditions to guarantee stability under switching. Compared to common Lyapunov function approaches, MLF provides greater flexibility for systems lacking a universal Lyapunov function, while accommodating broader classes of switching signals.

In general, constructing Lyapunov functions for arbitrary systems is a very challenging task. 
Analytical methods are only applicable to a restricted class of systems with favorable structural properties. 
While numerical approaches, such as sum-of-squares (SOS) programming \cite{papachristodoulou2002construction,tan2008stability,chen2023data}, offer a more general computational framework, they still face scalability limitations due to computational complexity in high-dimensional systems.
In recent years, Neural Lyapunov Functions (NLFs) have emerged as a powerful alternative for stability analysis \cite{chang2019neural, abate2020formal}.
By leveraging neural networks to approximate Lyapunov functions, NLFs can capture complex stability properties that traditional analytical methods often fail to address. 
These data-driven approaches enable stability verification for nonlinear systems with high-dimensional state spaces and even unknown dynamics. 
Moreover, NLF-based methods provide key advantages including fully-automated synthesis, scalability, and adaptability to diverse system configurations.

In this paper, we propose a new stability verification method for switched systems that combines the advantages of multiple Lyapunov functions   and neural Lyapunov functions. 
We consider a general state-dependent switching setting, where mode transitions are permitted within specific regions associated with each switching behavior.
Our approach introduces a Neural Multiple Lyapunov Function (NMLF) framework to rigorously verify stability in state-dependent switched systems. The framework employs a set of carefully designed loss functions along with a Counter-Example Guided Inductive Synthesis (CEGIS) scheme to train neural networks that satisfy MLF conditions. Through extensive simulations and theoretical analysis, we demonstrate the effectiveness of our method and its potential for practical applications in complex switched systems.

\subsection{Related Works}
As mentioned above, some recent works in the literature have explored the use of neural Lyapunov functions  for stability verification and controller synthesis of dynamic systems. 
For instance, \cite{chang2019neural, abate2020formal, richards2018lyapunov, 10560463, 10151866, dai2021lyapunov, https://doi.org/10.1002/rnc.5399, 9993006} has demonstrated how NLF can effectively verify stability while simultaneously maximizing the region of attraction (ROA). In contrast, \cite{10313502, jena2022distributedlearningneurallyapunov} focuses on reducing computational complexity while ensuring stability through the use of NLF. However, these works primarily focus on a non-switched nonlinear systems, and their applicability to hybrid systems, particularly region-based switching systems, remains limited. Our research aims to extend these efforts by applying NLF to the stability verification of hybrid systems with state constraints.

Recent studies have explored the integration of neural networks with multiple Lyapunov functions, including \cite{niu2019multiple,long2019multiple,yang2022adaptive}. However, these works primarily employ neural networks for adaptive controller synthesis rather than direct Lyapunov function approximation for stability verification. 
Notably, existing approaches do not specifically address the unique requirements of switched systems in their neural network training procedures. Our work addresses this gap by developing a tailored framework that combines neural networks with the MLF methodology, enabling direct approximation of Lyapunov functions for switched system stability verification.

\section{Preliminaries}\label{preliminary}
In this work, we consider the stability verification problem for switched systems in both  continuous-time setting
and  discrete-time setting.
Specifically, for the continuous-time setting, the system dynamic is given by:
\begin{equation}\label{eq:sys-cont}
    \dot{x} = f_i(x), \quad i \in Q = \{1, 2, \dots, N\},
\end{equation}
where 
\( x \in \mathbb{R}^n \) represents the system state, 
$Q $ is the set of $N$ system modes
and  
each \( f_i: \mathbb{R}^n \to \mathbb{R}^n \) is a vector field corresponding to the \( i \)-th mode of the system.
We assume that each function \( f_i(x) \) is locally Lipschitz continuous, which ensure the existence and uniqueness of solutions. 


We also consider the discrete-time setting, and the system dynamic is given by
\begin{equation}\label{eq:sys-disc}
    x_{k+1} = f_i(x_k), \quad i \in Q = \{1, 2, \dots, N\},
\end{equation}
where 
\( x_k \in \mathbb{R}^n \) denotes the system state at time step \( k \), 
and each \( f_i: \mathbb{R}^n \to \mathbb{R}^n \) is a mapping corresponding to the \( i \)-th mode. 

To analyze the stability of the switched systems, we first recall the notion of asymptotic stability.
\begin{definition}[Asymptotic Stability]\upshape
A switched system is said to be \emph{asymptotically stable} at an equilibrium point \(x^*\) in the region \(\mathcal{D}\) if, for every initial condition \(x(0) \in \mathcal{D}\), the corresponding solution (with \(t \geq 0\) for continuous-time systems or \(k \geq 0\) for discrete-time systems) satisfies
$\lim_{t\to\infty} x(t) = x^* \quad \text{(for continuous-time systems)}\nonumber$
 or 
$\lim_{k\to\infty} x_k = x^* \quad \text{(for discrete-time systems)}.\nonumber$
\end{definition}


In order to analyze stability of such switched systems, we employ  the tool of multiple Lyapunov functions (MLFs), which allow for stability analysis under arbitrary or constrained switching rules. Given a continuously differentiable function \( V: \mathbb{R}^n \to \mathbb{R} \), the Lie derivative of \( V \) along the vector field \( f_i(x) \) is defined as:
\begin{equation}
    L_{f_i} V (x) = \frac{d}{dt} V(x) \Big|_{\dot{x} = f_i(x)} = \nabla V(x)^\top f_i(x).
\end{equation}
Then the MLFs are formally defined as follows. 

\begin{definition}[\bf Multiple Lyapunov Functions]\upshape
Given a switched system 
either in the continuous-time setting of form \eqref{eq:sys-cont}
or in the discrete-time setting of form \eqref{eq:sys-disc}, 
we say   $\{V_i(x)\}_{i\in Q}$ is a multiple Lyapunov function  with respect to region \(\mathcal{D}\) and   equilibrium point \( x^* \), if the following conditions hold:  
 \begin{enumerate}
    \item \textbf{Mode-wise Lyapunov Function Properties:} 
    For each mode \( i \), we have
    $V_i(x^*) = 0 \quad \text{and} \quad V_i(x) > 0, \quad \forall x \in \mathcal{D} \setminus \{x^*\}.$
    Furthermore, in the continuous-time case, 
    \( V_i \) is continuously differentiable and its derivative along \( f_i(x) \) satisfies   
    $L_{f_i} V_i(x) = \nabla V_i(x)^\top f_i(x) < 0, \quad \forall x \in \mathcal{D} \setminus \{x^*\}.$   
    In the discrete-time case, the corresponding condition is
    $V_i(f_i(x)) - V_i(x) < 0, \quad \forall x \in \mathcal{D} \setminus \{x^*\},$
    and no differentiability assumption is required.
    \item 
    \textbf{Switching Decrease Properties:} 
    Whenever the system switches from mode \( i \) to mode \( j \), the multiple Lyapunov function  satisfy
    $V_j(x) < V_i(x)$.

\end{enumerate}
\end{definition}\upshape

Then we have the following theorem for MLFs \cite{664150}.

\begin{theorem}\upshape
For a switched system, if 
one can find  a MLF $\{V_i(x)\}_{i\in Q}$  with respect to   region \(\mathcal{D}\) and   equilibrium point \( x^* \), 
then the switched system is  asymptotically stable  in \(\mathcal{D}\). 
That is, for any initial state \( x(0) \in \mathcal{D} \) (or \( x_0 \in \mathcal{D} \) in the discrete case), the corresponding solution converges to \( x^* \) as \( t \to \infty \) (or as \( k \to \infty \) in the discrete-time case).
\end{theorem}

\begin{remark}\label{rmk1}
In some situations, the  mode-wise Lyapunov function properties are enforced only on 
\(
\mathcal{D} \setminus B_{\epsilon_b}(x^*),
\)
i.e., by excluding a ball of radius \(\epsilon_b > 0\) centered at the equilibrium \(x^*\). 
The in this case, one can conclude that every trajectory from an initial state \( x(0) \in \mathcal{D} \) (or \( x_0 \in \mathcal{D} \) in the discrete case) will ultimately enter and remain within the ball \(B_{\epsilon_b}(x^*)\). In other words, even though the Lyapunov conditions are not verified arbitrarily close to \(x^*\), the system is practically stable in the sense that its state converges to an \(\epsilon_b\)-neighborhood of the equilibrium. This result is particularly useful in numerical implementations, as it mitigates pathological issues such as arithmetic underflow near \(x^*\) while still guaranteeing a meaningful stability property.
\end{remark}

\section{Problem Formulation}
\label{sec:problem}
In the literature, the stability verification problem has been studied under various switching rules, such as arbitrary switching and language-constrained switching. Here, we consider a   class of \emph{state-dependent switching constraints}, where mode transitions are restricted to specific regions of the state space.

Formally, we consider a collection of pairwise disjoint switching regions \(\{\mathcal{D}_{ij}\}_{(i,j) \in Q \times Q}\), where each \(\mathcal{D}_{ij} \subseteq \mathcal{D}\).  
The switching rule is given by: 
When the system is in mode \(i\) and the state \(x\) enters \(\mathcal{D}_{ij}\), it may either:  
\begin{itemize}
\item Switch from mode \(i\) to \(j\), or
\item Remain in mode \(i\). 
\end{itemize}
The choice between staying or switching is nondeterministic whenever the switching condition is met. 
Our goal is to verify stability under all possible switching scenarios, which is formulated as follows. 

\begin{problem}[\bf Stability Verification under State-Dependent Switchings]
\label{prob:main}
Given a switched system 
either in the continuous-time setting of form \eqref{eq:sys-cont}
or in the discrete-time setting of form \eqref{eq:sys-disc}, with state-dependent switching constraints, 
design a computational framework to construct a multiple Lyapunov function  $\{V_i: \mathbb{R}^n \to \mathbb{R}\}_{i \in Q}$ satisfying the Mode-wise Lyapunov Properties enforced only on 
\(\mathcal{D} \setminus B_{\epsilon_b}(x^*)\) 
and switching decrease properties. 
\end{problem}

In the problem formulation above, we adopt a state-dependent switching framework motivated by both practical considerations and theoretical concerns: 
\begin{itemize}
\item 
First, the state-dependent switching framework  generalizes arbitrary switching. Particularly, by setting each $\mathcal{D}_{ij} = \mathcal{D}$, we recover arbitrary switching as a special case. However, real-world systems typically exhibit physical constraints that prevent truly arbitrary switching as mode transitions can only occur when specific state-dependent predicates are satisfied.
\item 
As our analysis will demonstrate, arbitrary switching may induce \emph{Lyapunov mode collapse} in our learning-based framework. This phenomenon occurs when transition compatibility conditions ($V_j(x) < V_i(x)$) must hold across unconstrained switching regions, forcing distinct Lyapunov functions $\{V_i\}$ to adopt similar geometric structures. The conflicting gradient directions during neural network training effectively paralyze the learning process. Our Neural MLF framework avoids this degeneracy by incorporating state-dependent switching constraints, enabling stable learning of distinct Lyapunov functions while maintaining training efficiency.
\end{itemize}

\section{Methodology}
\label{sec:methodology}

In this work, we propose a neural framework for synthesizing Lyapunov functions that certify the stability of state-constrained switching systems. The problem is addressed in two parts: (i) constructing mode-wise candidate Lyapunov functions with tailored loss terms to enforce mode-specific stability properties, and (ii) imposing switching decrease conditions to ensure stability across mode transitions. To guarantee these requirements, we employ a Counter-Example Guided Inductive Synthesis (CEGIS) framework, where an SMT solver verifies candidates and provides counterexamples for iterative refinement. The methodology is detailed in the following subsections.

\subsection{Mode-Wise NLFs and Mode-Wise Losses}
For each mode $i$, we construct a candidate Lyapunov function of the form
\begin{equation}\label{eq:NLF}
  V_{i,\theta}(x) = v_{i,\theta}(x) - v_{i,\theta}(0),
\end{equation}
where $v_{i,\theta}(\cdot)$ is a multilayer perceptron (MLP) with parameter vector $\theta$. This formulation guarantees that $V_{i,\theta}(0)=0$, a necessary property for any Lyapunov function candidate.

To enforce that the candidate function $V_{i,\theta}(x)$ satisfies the remaining \textbf{mode-wise Lyapunov function properties} over the entire working domain $\mathcal{D}$, we introduce the following loss function.

\begin{definition}[\bf Mode-Wise Loss Functions]\label{def:loss} 
Let $\rho$ be a sampling distribution defined on $\mathcal{D}$.
The mode-wise loss function for mode \( i \) is defined by:
\begin{itemize}
    \item 
    {Continuous-time case:}  
\begin{equation}\label{eq:loss_expectation_ct}
\begin{split}
\mathcal{L}_{\rho}^{\text{ct}}(\theta;i) =\, & \mathbb{E}_{x\sim\rho(\mathcal{D})}\Bigl[
      \max\Bigl\{ 0,\; L_{f_i}V_{i,\theta}(x) + \epsilon \Bigr\}
    \Bigr] \\
& + \alpha\, \mathbb{E}_{x\sim\rho(\mathcal{D})}\Bigl[
      \max\Bigl\{ 0,\; \epsilon - V_{i,\theta}(x) \Bigr\}
    \Bigr].
\end{split}
\end{equation}
    \item 
   {Discrete-time case:}  
\begin{align}\label{eq:loss_expectation_dt}\!\!\!\!\!\!\!\!\!\!\!\!\!
\mathcal{L}_{\rho}^{\text{dt}}(\theta;i) =\, & \mathbb{E}_{x_k\sim\rho(\mathcal{D})}\Bigl[
      \max\Bigl\{ 0,\; V_{i,\theta}(f_i(x_{k})) - V_{i,\theta}(x_k) + \epsilon \Bigr\}
    \Bigr] \nonumber\\
& + \alpha\, \mathbb{E}_{x_k\sim\rho(\mathcal{D})}\Bigl[
      \max\Bigl\{ 0,\; \epsilon - V_{i,\theta}(x_k) \Bigr\}
    \Bigr].
\end{align}
\end{itemize}
\end{definition}

Note that, the above defined loss functions depend on the distribution $\rho$ 
and the expection is difficult to compute in general. 
In practice,  the loss is computed in  a data-driven (empirical)  fashion by  by sampling \( N \) points \(\{x_j\}_{j=1}^{N}\) from \(\mathcal{D}\) according to \(\rho\), leading to the following \textbf{empirical loss functions} defined as follows:
\begin{itemize}
    \item 
    {Continuous-time case:}  
\begin{align}\label{eq:loss_empirical_ct} 
\mathcal{L}_{N,\rho}^{\text{ct}}(\theta;i) =\, & \frac{1}{N}\sum_{j=1}^N\Bigl[
      \max\Bigl\{ 0,\; L_{f_i}V_{i,\theta}(x_j) + \epsilon \Bigr\}
    \Bigr] \\
& + \alpha\, \frac{1}{N}\sum_{j=1}^N\Bigl[
      \max\Bigl\{ 0,\; \epsilon - V_{i,\theta}(x_j) \Bigr\}
    \Bigr]. \nonumber
\end{align}
\item 
{Discrete-time case:}  
\begin{align}\label{eq:loss_empirical_dt}
\!\!\!\!\!\!\!\! 
\mathcal{L}_{N,\rho}^{\text{dt}}(\theta;i) =\, & \frac{1}{N}\sum_{j=1}^N\Bigl[
      \max\Bigl\{ 0,\; V_{i,\theta}(f_i(x_{j})) - V_{i,\theta}(x_j)\! + \!\epsilon \Bigr\}\!\!
    \Bigr] \nonumber\\
& + \alpha\, \frac{1}{N}\sum_{j=1}^N\Bigl[
      \max\Bigl\{ 0,\; \epsilon - V_{i,\theta}(x_j) \Bigr\}
    \Bigr].
\end{align}
\end{itemize}

Intuitively, minimizing either \( \mathcal{L}_{\rho}^{\text{ct}}(\theta;i) \) or its empirical counterpart \( \mathcal{L}_{N,\rho}^{\text{ct}}(\theta;i) \) ensures that the candidate Lyapunov function satisfies $L_{f_i}V_{i,\theta}(x) \le -\epsilon \quad \text{and} \quad V_{i,\theta}(x) \ge \epsilon$,

for all \( x\in\mathcal{D} \) (except at the equilibrium \( x=0 \)), thereby enforcing the desired Lyapunov conditions for mode \( i \).  
Similarly, minimizing \( \mathcal{L}_{\rho}^{\text{dt}}(\theta;i) \) or \( \mathcal{L}_{N,\rho}^{\text{dt}}(\theta;i) \) ensures the discrete-time system satisfies $V_{i,\theta}(f_i(x_{k})) \le V_{i,\theta}(x_k) - \epsilon \quad \text{and} \quad V_{i,\theta}(x_k) \ge \epsilon$, for all \( x_k\in\mathcal{D} \), thus enforcing the Lyapunov conditions in the discrete-time case.

\subsection{Switching Losses}
In  region \(\mathcal{D}_{ij}\), where the system is allowed to switch from mode \(i\) to mode \(j\), the desired property is that
\[
V_{i,\theta}(x) > V_{j,\theta}(x), \quad \forall\, x\in \mathcal{D}_{ij}.
\]
To enforce this requirement during the training process, we introduce the following loss function.

\begin{definition}[\bf Switching Loss Functions]\label{def:switching_loss}
Let $\rho$ be a sampling distribution defined on $\mathcal{D}$
and $\epsilon > 0$ be a prescribed margin parameter.
For a given margin \(\epsilon>0\), the switching loss is defined by:
\begin{itemize}
    \item 
     {Continuous-time case:}
    \begin{equation}\label{eq:switching_loss_ct_dist}
        \!\!\!\!\!\!\!\!
        \mathcal{L}_{\rho}^{(i,j),\text{ct}}(\theta)
      = \mathbb{E}_{x\sim\rho(\mathcal{D}_{ij})}\Bigl[
          \max\Bigl\{ 0,\; V_{j,\theta}(x) - V_{i,\theta}(x) + \epsilon \Bigr\}
        \Bigr].
    \end{equation}
    \item 
    {Discrete-time case:}
    \begin{equation}\label{eq:switching_loss_dt_dist}
        \!\!\!\!\!\!\!\!\!\!\!\!\!\!
        \mathcal{L}_{\rho}^{(i,j),\text{dt}}(\theta)
      \!= \!\mathbb{E}_{x_k\sim\rho(\mathcal{D}_{ij})}\Bigl[
          \max\Bigl\{ 0,\!V_{j,\theta}(f_j(x_k)) \!-\! V_{i,\theta}(x_k) \!+ \!\epsilon \Bigr\}
        \Bigr].
    \end{equation}
\end{itemize}
\end{definition}
  
Similar to \ref{def:loss}, we can define a empirical version of switch loss functions by sampling $N$ points \(\{x_k\}_{k=1}^{N}\) from \(\mathcal{D}_{ij}\) according to \(\rho\).
\begin{itemize}
    \item 
    {Continuous-time case:}
    \begin{equation}\label{eq:switching_loss_ct_sample}
        \mathcal{L}_{N,\rho}^{(i,j),\text{ct}}(\theta)
      = \frac{1}{N}\sum_{k=1}^{N} \max\Bigl\{ 0,\; V_{j,\theta}(x_k) - V_{i,\theta}(x_k) + \epsilon \Bigr\}.
    \end{equation}
   \item
    {Discrete-time case:}
    \begin{equation}\label{eq:switching_loss_dt_sample}
           \!\!\!\!\!\!\!\!\!\!\!\!  \mathcal{L}_{N,\rho}^{(i,j),\text{dt}}(\theta)
      = \frac{1}{N}\sum_{k=1}^{N} \max\Bigl\{ 0,\; V_{j,\theta}(f_j(x_k)) - V_{i,\theta}(x_k) + \epsilon \Bigr\}.
    \end{equation}
\end{itemize}

\subsection{Verification  and Counter-Example Generations}
\label{sec:verification}
In our work, we adopt a Counter-Example Guided Inductive Synthesis (CEGIS) framework to ensure that the candidate Lyapunov functions satisfy the desired properties. 
Let \(x^*\) denote the equilibrium point. 
To avoid pathological numerical issues such as arithmetic underflow near \(x^*\), we exclude a ball of radius \(\epsilon_b > 0\) centered at \(x^*\). Specifically, we define the verification region as
$\mathcal{D}_V = \{ x \in \mathcal{D} \mid \|x-x^*\| \ge \epsilon_b \}.$

Within \(\mathcal{D}_V\), the candidate Lyapunov functions must satisfy the mode-wise Lyapunov properties and the switching decrease properties. We encapsulate the failure of these properties through the violation predicate \(\Phi(x)\). For \(x \in \mathcal{D}_V\), we define:

\textbf{Continuous-time case:}
\begin{equation}
\begin{split}
\Phi(x) \coloneqq {} & \Biggl( \bigvee_{i \in \mathcal{I}} \Bigl\{ V_i(x) \le 0 \vee L_{f_i}V_i(x) \ge 0 \Bigr\} \Biggr) \\
& \vee \Biggl( \bigvee_{(i,j)} \Bigl[ x \in \mathcal{D}_{ij} \wedge \{ V_j(x) \ge V_i(x) \} \Bigr] \Biggr).
\end{split}
\end{equation}

\textbf{Discrete-time case:}
\begin{equation}
\begin{split}
\Phi(x) \coloneqq {} & \Biggl( \bigvee_{i \in \mathcal{I}} \Bigl\{ V_i(x) \le 0 \vee \Bigl[ V_i(f_i(x))-V_i(x) \ge 0 \Bigr] \Bigr\} \Biggr) \\
& \vee \Biggl( \bigvee_{(i,j)} \Bigl[ x \in \mathcal{D}_{ij} \wedge \{ V_j(x) \ge V_i(x) \} \Bigr] \Biggr).
\end{split}
\end{equation}

That is, for every \(x \in \mathcal{D}_V\), if there exists some mode \(i\) for which either \(V_i(x) \le 0\) or the corresponding Lyapunov decrease condition (i.e., \(L_{f_i}V_i(x) < 0\) in the continuous case or \(V_i(f_i(x))-V_i(x) < 0\) in the discrete case) is violated, then \(\Phi(x)\) evaluates to true. In addition, for any switching region \(\mathcal{D}_{ij}\), if there exists \(x \in \mathcal{D}_{ij}\) such that \(V_j(x) \ge V_i(x)\), then \(\Phi(x)\) is true.

An SMT solver is employed to search for any \(x_e \in \mathcal{D}_V\) that satisfies \(\Phi(x_e)\). If such a counter-example \(x_e\) is found, it is incorporated into the training set for further refinement of the candidate Lyapunov functions. Otherwise, if no counter-example exists, the candidate is deemed to satisfy the required properties. This verification step is executed after each iteration within our CEGIS loop, ensuring that the candidate Lyapunov functions conform to the mode-wise and switching decrease properties (with the discrete-time case enforcing \(x_{k+1}=f_j(x_k)\) upon mode switching).

As discussed in \cite{chang2019neural}, 
the main purpose of excluding the region \(B_{\epsilon_b}(x^*)\) from the verification process is to mitigate  numerical sensitivity issues (e.g., arithmetic underflow) while preserving the properties of the Lyapunov level sets and the regions of attraction outside this excluded ball.

\subsection{Training Algorithm}
Combining the two components described above, 
we define the overall   empirical loss function i  as follows. For the continuous--time case, the overall loss is given by
\begin{equation}\label{eq:overall_loss_ct}
  \mathcal{L}^{\text{ct}}(\theta)
  = \sum_{i} \mathcal{L}_{N,\rho}^{\text{ct}}(\theta;i)
    + \beta\sum_{(i,j)} \mathcal{L}_{N_{ij},\rho}^{(i,j),\text{ct}}(\theta),
\end{equation}
where \(\mathcal{L}_{N,\rho}^{\text{ct}}(\theta;i)\) is the empirical loss for mode \(i\) (see Equation~\eqref{eq:loss_empirical_ct} in the previous section) and \(\mathcal{L}_{N_{ij},\rho}^{(i,j),\text{ct}}(\theta)\) is the empirical switching loss for the mode pair \((i,j)\) defined over \(\mathcal{D}_{ij}\). In the discrete--time case, the overall loss is defined analogously:
\begin{equation}\label{eq:overall_loss_dt}
  \mathcal{L}^{\text{dt}}(\theta)
  = \sum_{i} \mathcal{L}_{N,\rho}^{\text{dt}}(\theta;i)
    + \beta\sum_{(i,j)} \mathcal{L}_{N_{ij},\rho}^{(i,j),\text{dt}}(\theta),
\end{equation}
with the additional requirement that when switching from mode \(i\) to mode \(j\) we have \(x_{k+1}=f_j(x_k)\).

The overall training procedure is summarized in Algorithm~\ref{alg:training}. In brief, we sample states from \(\mathcal{D}\) and the switching regions \(\mathcal{D}_{ij}\) according to the distribution \(\rho\), compute the overall empirical loss \(\mathcal{L}^{\ast}(\theta)\), and update the network parameters \(\theta\) via gradient descent. Subsequently, as detailed in Section~\ref{sec:verification}, an SMT solver is employed to verify that the candidate Lyapunov functions satisfy both the mode-wise Lyapunov properties and the switching decrease conditions over the verification region 
$\mathcal{D}_V = \{ x \in \mathcal{D} \mid \|x-x^*\| \ge \epsilon_b \},$
(with \(x_{k+1}=f_j(x_k)\) enforced in the discrete-time case). If a counterexample is found, it is incorporated into the training set and the process is repeated until no counterexamples exist within the specified tolerance.

\begin{algorithm}[t]
\caption{Training Algorithm for Mode-wise Neural Lyapunov Functions (CEGIS)}\label{alg:training}
\begin{algorithmic}[1]
\State \textbf{Input:} $\theta_0$, sample sets $S\subset\mathcal{D}$, $\{S_{ij}\}$ for switching regions $\mathcal{D}_{ij}$, tolerance $\delta>0$
\State $\theta \gets \theta_0$
\Repeat
    \State For each mode $i$, compute $\mathcal{L}_{N,\rho}^{*}(\theta;i)$ on $S$, $*\!\in\!\{\mathrm{ct},\mathrm{dt}\}$
    \State For each mode pair $(i,j)$ \emph{with switching region $\mathcal{D}_{ij}$}, compute $\mathcal{L}_{N_{ij},\rho}^{(i,j),*}(\theta)$ on $S_{ij}$
    \State Update $\theta \gets \theta - \eta\, \nabla_\theta \big[\sum_i \mathcal{L}_{N,\rho}^{*}(\theta;i) 
          + \beta\sum_{(i,j)} \mathcal{L}_{N_{ij},\rho}^{(i,j),*}(\theta)\big]$
    \State \textbf{Verification:} use SMT solver (Sec.~\ref{sec:verification}) to check Lyapunov and switching conditions over $\mathcal{D}_V$; if counterexamples $X^*$ found, add them to $S$ or $S_{ij}$ accordingly
\Until{no counterexamples in $\mathcal{D}_V$ within $\delta$}
\State \textbf{Output:} $\theta$
\end{algorithmic}
\end{algorithm}

This training algorithm integrates the empirical loss from both the individual mode conditions and the switching requirements, followed by a verification step via an SMT solver. The incorporation of CEGIS ensures that the training progressively refines the candidate Lyapunov functions until the desired conditions are satisfied. According to remark~\ref{rmk1}, if we can obtain a set of neural multiple Lyapunov functions from algorithm~\ref{alg:training}, then for any initial state \( x(0) \in \mathcal{D} \) (or \( x_0 \in \mathcal{D} \) in the discrete case), the corresponding solution will finally enter and remain within $\mathcal{B}_{\epsilon_b}$.

\section{Experimental Results}
\label{others}

We demonstrate the correctness and efficiency of Algorithm~\ref{alg:training} across a variety of experiments. All experiments were conducted on a single NVIDIA GeForce RTX 3090 GPU. Furthermore, in the verification step, we employ the SMT solver \emph{dReal} to solve nonlinear, non-polynomial disjunctive constraint systems. All codes are available at \url{https://github.com/JunyueHuang/Neural_MLF}.

\subsection{Pendulum Systems}
In the first experiment, we consider a nonlinear continuous-time pendulum system operating on a two-dimensional state \(x = [s,\,v]^\top\) within the working domain
$\mathcal{D} = \{x \in \mathbb{R}^2 : \|x\| \leq 3\}.$
The system dynamic is  given by
$\dot{s} = v,\quad \dot{v} = \frac{-m\,G\,L\,\sin(s) - b\,v}{m\,L^2},$
where \(s\) and \(v\) denote the angular position and velocity, respectively, and \(m\), \(G\), and \(L\) represent the mass, gravitational constant, and pendulum length. 

\begin{figure}[t]\label{fig:aa} 
\centering
	\subfigure[System model.]
	{\label{fig:pic1}
		\includegraphics[height=4cm]{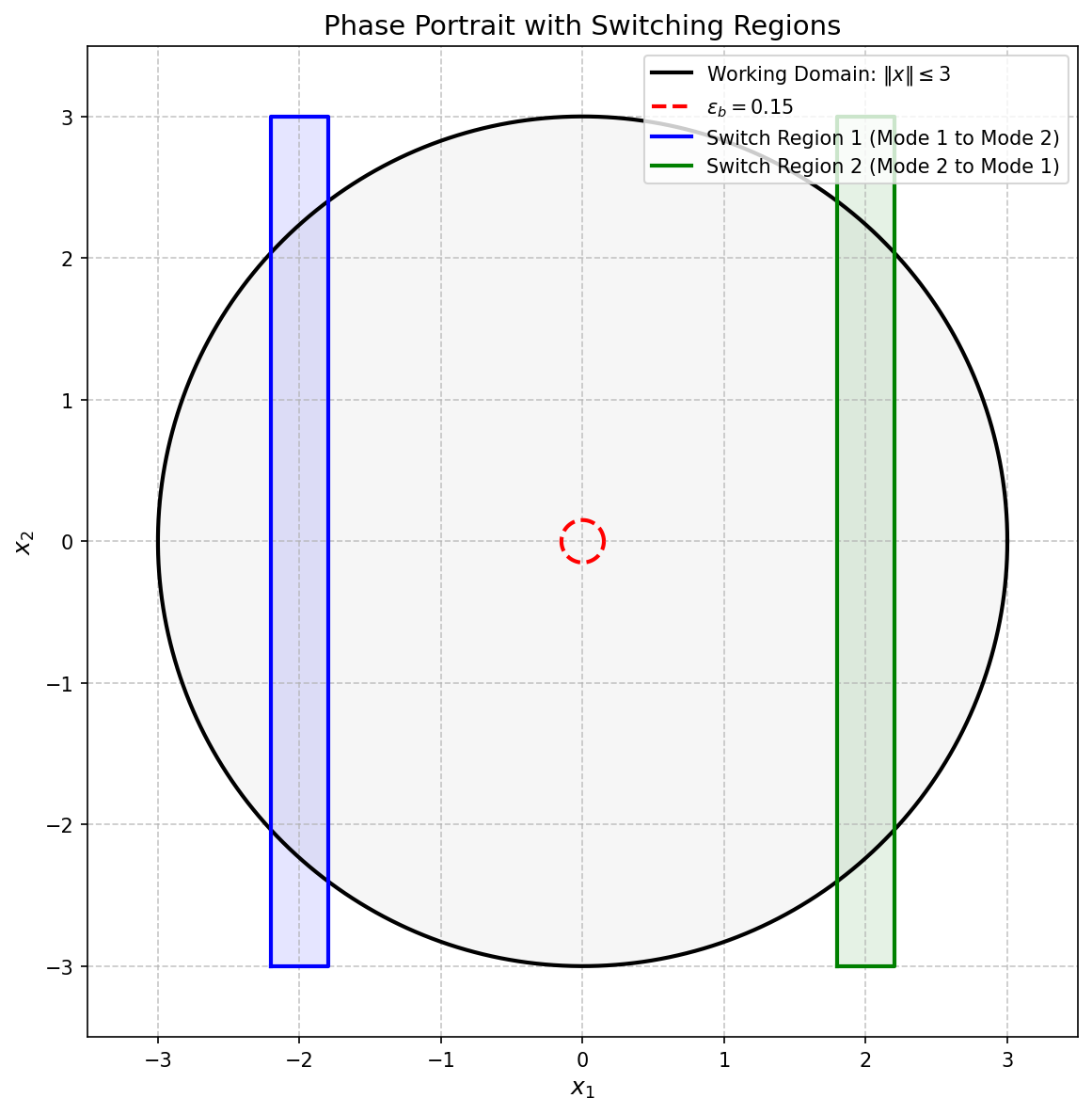} } 
	\subfigure[Neural MLF.]
	{\label{fig:pic2} 
		\includegraphics[height=4cm]{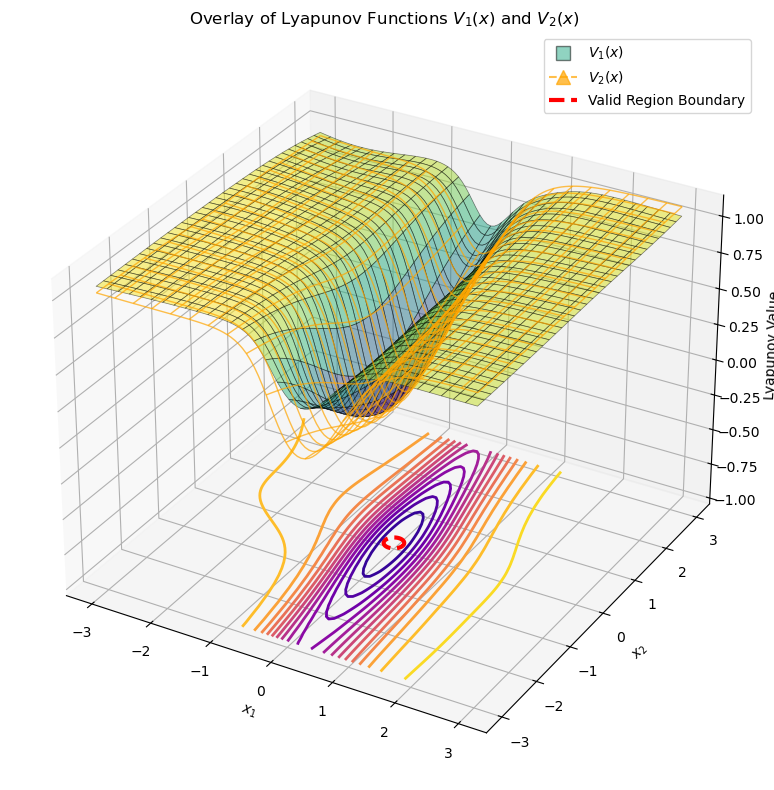} }
	\caption{Experiment in Section V.A.}\label{fig:a}
\end{figure}

We consider two operating modes of the system:
\begin{itemize}
    \item 
    \textbf{mode~1} with damping \(b=0.1\); and 
    \item 
    \textbf{mode~2} with damping \(b=0.3\). 
\end{itemize}  
Switching regions are defined as: from mode~1 to mode~2,
we have 
\(\mathcal{D}_{12} = [-2.2,-1.8,-3,3]\); and from mode~2 to mode~1, 
we have 
\(\mathcal{D}_{21} = [1.8,2.2,-3,3]\) (see Fig.~\ref{fig:pic1}).
Furthermore, a ball of radius \(\epsilon_b = 0.15\) centered at the equilibrium is excluded during verification to mitigate numerical issues. 

By applying our training algorithm,  we successfully   synthesize a  multiple neural Lyapunov functions. 
The values of these two functions are shown graphically in Fig.~\ref{fig:pic2}, which  satisfy the prescribed conditions and therefore, verify the stability of the system. 

\subsection{No Common Lyapunov Function Exists} 
\begin{figure}[t]\label{fig:bb} 
\centering
	\subfigure[Instability for arbitrary switching]
	{\label{fig:pic3}
		\includegraphics[height=4cm]{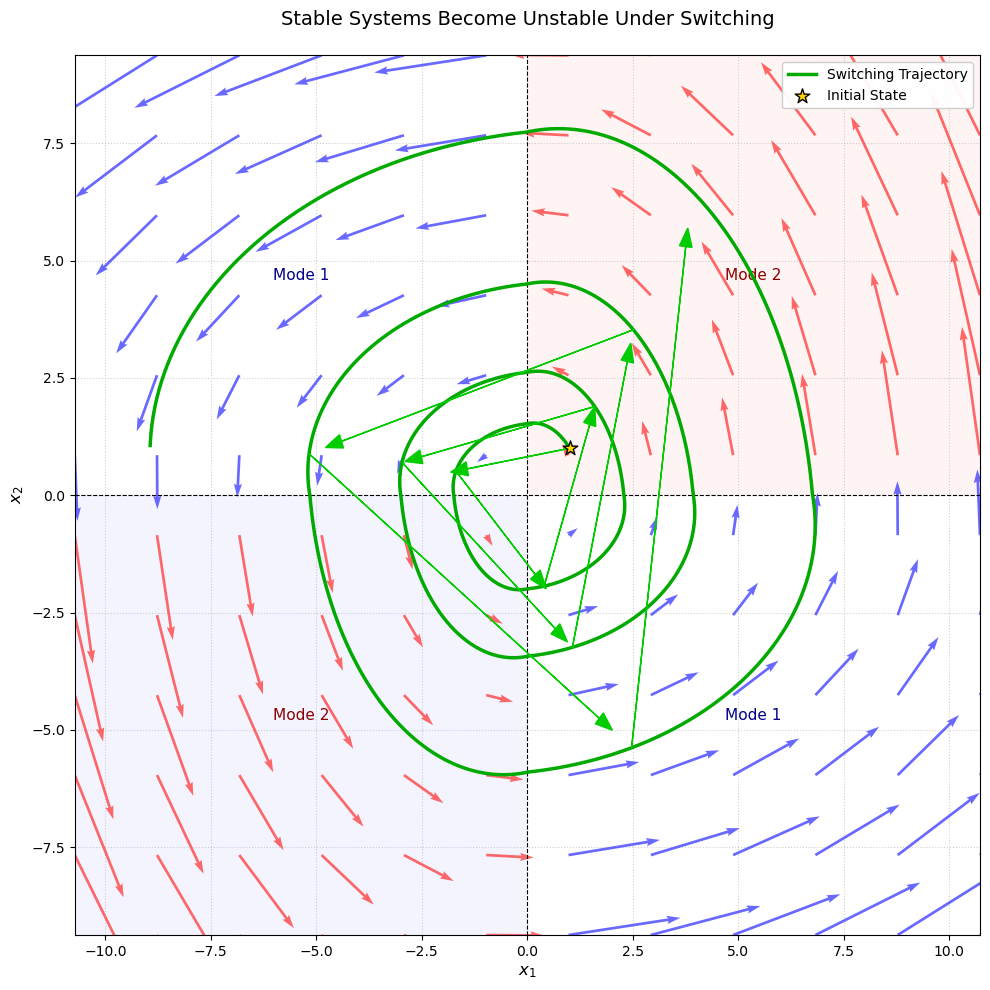} } 
	\subfigure[Neural MLF.]
	{\label{fig:pic4} 
		\includegraphics[height=4cm]{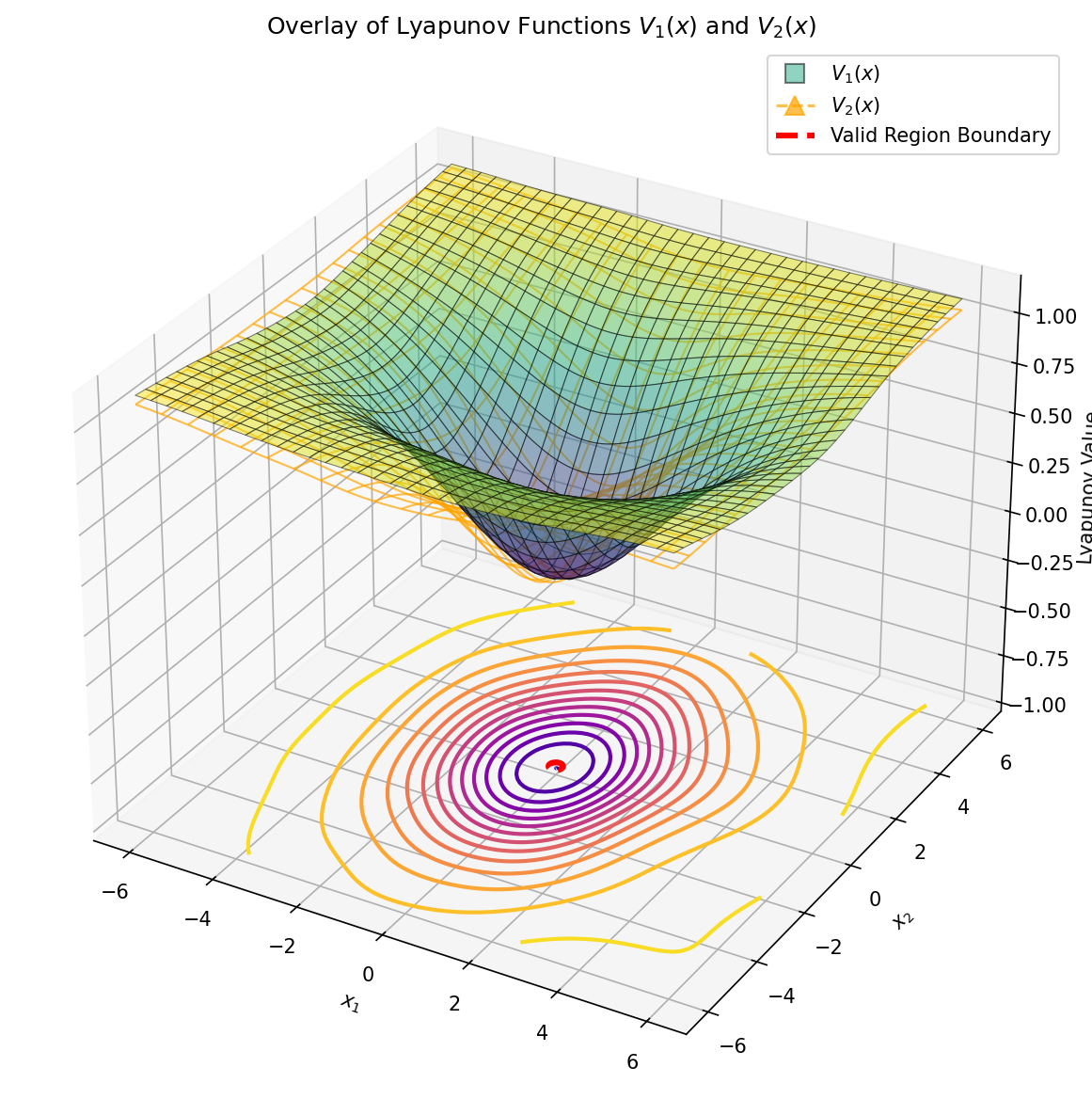} }
	\caption{Experiment in Section V.B.}\label{fig:b}
\end{figure}

In the second experiment, we consider two linear systems with dynamics 
$\dot{x} = A_i x,\quad i=1,2,$

where 
\[
A_1 = \begin{bmatrix} -0.2 & -2.0 \\ 1.0 & -0.2 \end{bmatrix},\quad
A_2 = \begin{bmatrix} -0.2 & -1.0 \\ 2.0 & -0.2 \end{bmatrix}.
\]
The working domain is defined as 
$
\mathcal{D} = \{x \in \mathbb{R}^2 \mid \|x\|\leq 6\},
$
with a circular region of radius \(\epsilon_b=0.5\) (centered at the origin) excluded. 
Note that, under arbitrary switching between modes without state-constraint, 
the overall system becomes unstable (see Fig.~\ref{fig:pic3}), indicating that no common Lyapunov function exists.

Here, we further consider a state-dependent switching setting defined by: 
\begin{itemize}
    \item 
    from mode~1 to mode~2, the switching region is  \(\mathcal{D}_{12} = [-0.2,0.2,-2.2,-1.8]\); and 
    \item 
    from mode~2 to mode~1, the switching region is  \(\mathcal{D}_{21} = [1.8,2.2,-0.2,0.2]\).
\end{itemize}
By applying our training algorithm,  we can still synthesize a  multiple neural Lyapunov function as shown graphically in Fig.~\ref{fig:pic4}. 
Therefore, we can claim that this system, which is not stable under arbitrary switching, is stable under the given state-dependent switching setting.

\subsection{Multi-Mode-Switching} 
Finally, we consider experiments for switched systems with more than two modes. 
\subsubsection{Discrete-Time Case}
We consider a switched system with dynamics
$
\dot{x} = A_i x,\quad i=1,2,3,
$
where 
\[
A_1 = \begin{bmatrix} 0.6 & -0.1 \\ 0.2 & 0.8 \end{bmatrix},\ 
A_2 = \begin{bmatrix} 0.6 & 0.1 \\ 0 & 0.6 \end{bmatrix},\ 
A_3 = \begin{bmatrix} 0.7 & 0.1 \\ -0.3 & 0.7 \end{bmatrix}.
\]
The working domain is defined as 
$\mathcal{D} = \{x \in \mathbb{R}^2 \mid \|x\|\leq 6\},$
with a circular region of radius \(\epsilon_b=0.5\) (centered at the origin) excluded. 
The switching regions are defined as:
\begin{itemize}
    \item 
    from mode~1 to mode~2,   \(\mathcal{D}_{12} = [-2.5,-1.5,-0.5,0.5]\); 
    \item 
    from mode~2 to mode~3,    \(\mathcal{D}_{23} = [1.5,2.5,-0.5,0.5] \); and 
    \item 
    from mode~3 to mode~1,    \(\mathcal{D}_{31} = [-0.5,0.5,-2.5,-1.5]\).
\end{itemize}
The switching yields overall stable behavior 
as our learning algorithm successfully finds e a  multiple neural Lyapunov function as shown graphically in Fig.~\ref{fig:pic6}.

\subsubsection{Continuous Case}
\begin{figure}[t]\label{fig:cc} 
\centering
	\subfigure[Neural MLF for discrete  case.]
	{\label{fig:pic6}
		\includegraphics[height=4cm]{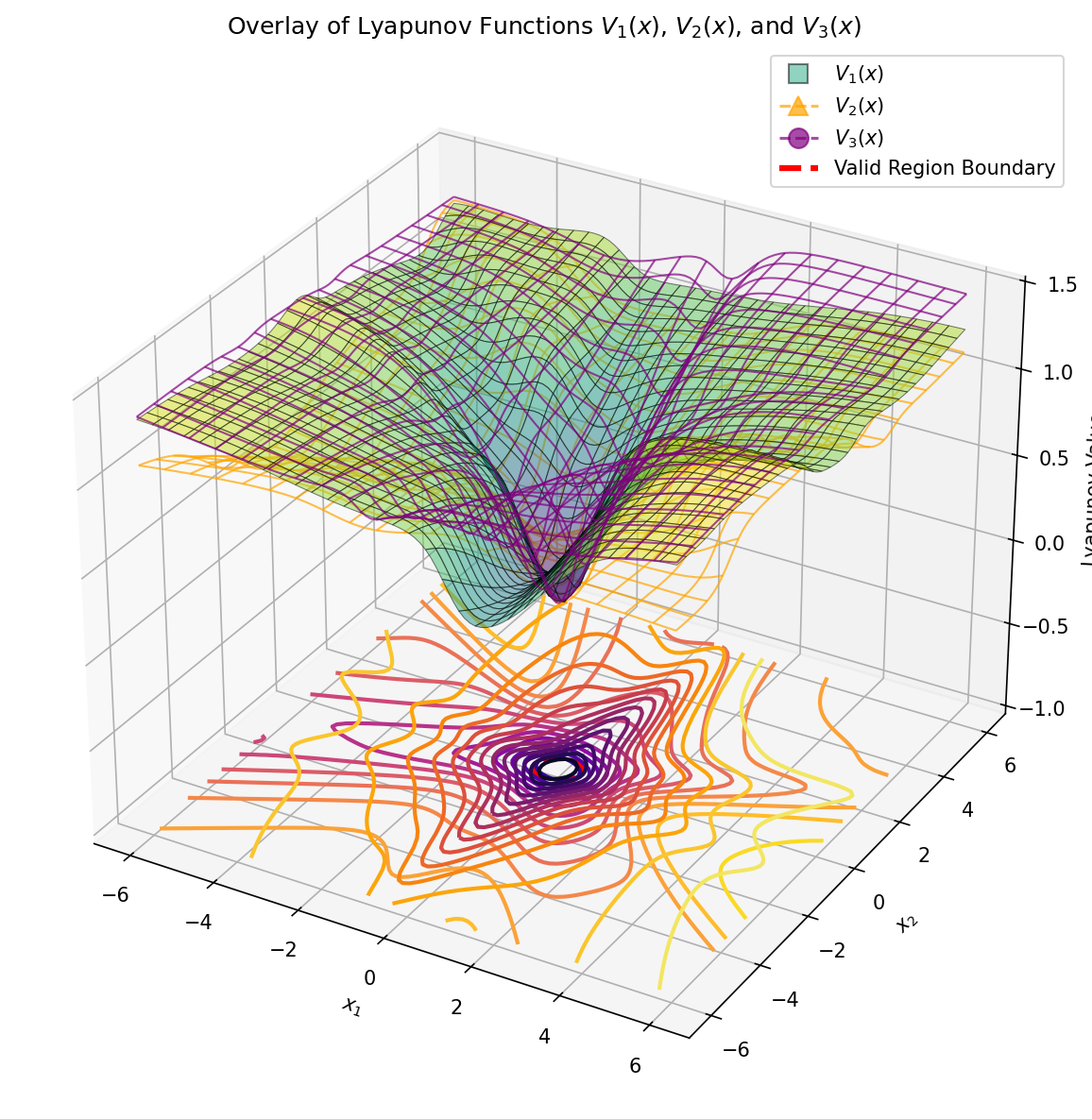} } 
	\subfigure[Neural MLF for continuous  case.]
	{\label{fig:pic7} 
		\includegraphics[height=4cm]{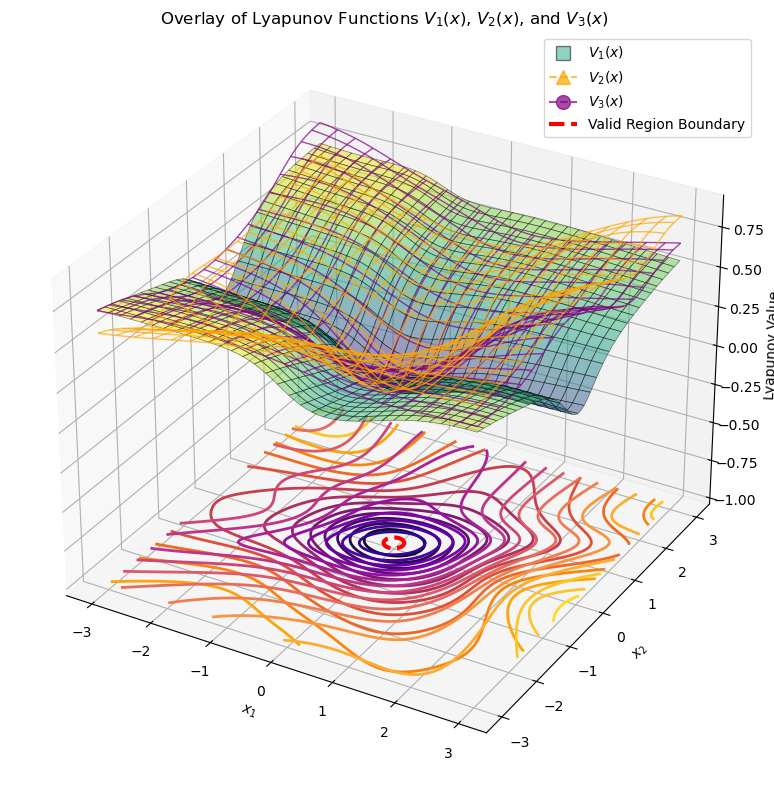} }
	\caption{Experiment in Section V.C.}\label{fig:c}
\end{figure}

Finally, we  consider a nonlinear system with three switching modes defined as:
\[
\dot{x} = f(x) = 
\begin{cases}
[-s + 2s^2v,\ -v]^\top       & (\text{Mode 1}) \\
[-s,\ -2v + 0.1sv^2]^\top   & (\text{Mode 2}) \\
[-3s - 0.1sv^3,\ -v]^\top   & (\text{Mode 3})
\end{cases}
\]
where \( x = [s,\,v]^\top \in \mathcal{D} = \{x \in \mathbb{R}^2 : \|x\| \leq 3\} \). 
The switching regions are defined as:
\begin{itemize}
    \item 
    from mode~1 to mode~2,   \(\mathcal{D}_{12} = [-0.5,0.5,1.5,2.5]\); 
    \item 
    from mode~2 to mode~3,    \(\mathcal{D}_{23} = [1.5,2.5,-0.5,0.5] \); and 
    \item 
    from mode~3 to mode~1,    \(\mathcal{D}_{31} = [-0.5,0.5,-2.5,-1.5]\).
\end{itemize}
As shown in Fig.~\ref{fig:pic7}, excluding a circular region \( \epsilon_b = 0.15 \) around the origin, the switching yields overall stable behavior.

\section{Conclusion}
This paper presented a neural multiple Lyapunov function (NMLF) framework for stability analysis of switched systems with state-dependent switching, combining the theoretical guarantees of multiple Lyapunov functions with the approximation power of neural networks. We developed a practical training algorithm and validated its effectiveness through case studies.
Future work includes extending the framework to broader classes of switched systems, such as those with language-constrained switching, and adapting it for control synthesis to jointly design control laws and switching strategies for stabilization.

\bibliographystyle{ieeetr}
\bibliography{MLF} 

\clearpage

\end{document}